\documentclass[a4paper]{llncs}
\usepackage{graphicx}
\usepackage{latexsym}
\usepackage{float}
\usepackage{amssymb}
\usepackage{amstext}
\usepackage{amsmath}
\usepackage{multirow}
\usepackage{xspace}
\usepackage{subfigure}
\usepackage{cite}
\usepackage[pdftex]{hyperref}

\newcommand{\df}{\ensuremath{\mathrel{:=}}}

\newcommand{\logAnd}{\mathrel{\wedge}}

\RequirePackage{dsfont}
\newcommand{\bag}{\mathds{B}}    

\newcommand{\trace}[1]{\langle #1 \rangle}

\title{Extracting and Pre-Processing Event Logs}

\author{
Dirk Fahland
} 

\institute{
Eindhoven University of Technology, The Netherlands\\
\email{d.fahland@tue.nl}
} 

\begin{document}
\maketitle
\pagestyle{headings}

\begin{abstract}
Event data is the basis for all process mining analysis. Most process mining techniques assume their input to be an \emph{event log}. However, event data is rarely recorded in an event log format, but has to be \emph{extracted} from raw data. Event log extraction itself is an act of \emph{modeling} as the analyst has to consciously choose which features of the raw data are used for describing which behavior of which entities. Being aware of these choices and subtle but important differences in concepts such as trace, case, activity, event, table, and log is crucial for mastering advanced process mining analyses.

This text provides fundamental concepts and formalizations and discusses design decisions in event log extraction from a raw event table and for event log pre-processing. It is intended as study material for an advanced lecture in a process mining course.
\end{abstract}

\section{Event Data}\label{sec:event:event_data}

Event data is the basis for all process mining analysis. Most process mining techniques assume that their input is in the form of a \emph{simple event log} such as the following:
$$
\begin{array}{rl}
L = [   & \trace{A,B,C,D}^{10},\\
        & \trace{A,C,B,D}^5,\\
        & \trace{A,B,A,D}^3,\\
        & \trace{A,E,D}^1 ].
\end{array}
$$
This simple event log is defined over an \emph{alphabet} $\Sigma = \{A,B,C,D,E\}$ which is a \emph{set} of activity names that have been observed. Each $a \in \Sigma$ is the name of an \emph{activity}, i.e., a specific action that can be executed or observed. For now, we consider each activity name as ``atomic''\,---\,later in this chapter we will see that activities themselves can have some ``structure'' themselves.

A \emph{trace} $\sigma \in \Sigma^*$ is a finite sequence of activities\footnote{Recall that the star ${}^*$ after $\Sigma$ is the \emph{Kleene star} which we use when constructing the set of all possible finite sequences over the elements of set $\Sigma$.}. It describes that this sequence of activities had been observed at some point in the past. Each occurrence of an activity in a trace $\sigma$ is called an \emph{event}. For example, the trace $\trace{A,B,A,D}$ describes we first observed $A$, then $B$ followed by another occurrence of $A$, and finally we observed $D$.

\begin{exercise}
Which other traces can be built from $\Sigma = \{A,B,C,D,E\}$ that are not in $L$?
\end{exercise}

A \emph{simple event log} $L \in \bag(\Sigma^*)$ is a multiset\footnote{A multi-set is also called a \emph{bag}, which explains the symbol $\bag$ we use for constructing the multiset over $\Sigma^*$. Recall that a multiset can contain the same element $\sigma \in \Sigma^*$ multiple times.} of traces describing that various traces that have been observed and \emph{how often} each trace has been observed, e.g., $\trace{A,B,A,D}^3$ was observed $3$ times.

\begin{exercise}
Why do we only study finite sequences of activities when analyzing event logs (recorded historic executions) of processes?
\end{exercise}

However, event data is \emph{not} recorded in this form in practice. First of all, an event records multiple attributes, not just the name of an activity. Secondly, event data is recorded as it occurs, and thus never grouped into traces or event logs.

A central part of process mining comprises actually obtaining event data from various data sources, and transforming it into an event log. We will see that both steps are non-trivial and allow for many choices. After the event log has been created, it rarely has sufficient quality to be used for any process mining analysis. Consequently, we have to pre-process the event log.

In the following, we first introduce a generic event data model and the notion of an event table in Sect.~\ref{sec:event:event_table}. Then, we explain in Sect.~\ref{sec:event:structured_log} how to extract structured event logs from such an event table and kinds of choices that can be made. We then introduce in Sect.~\ref{sec:event:classifiers} the notion of event classifiers required to turn structured event logs into the simple event logs explained above. We introduce the three central pre-processing operations on structured event logs in Sect.~\ref{sec:event:pre-processing}.

\section{Events and Event Table}\label{sec:event:event_table}

The most common direct or ``raw'' logging format for events is an \emph{event table} or \emph{event stream} as shown in Table~\ref{tab:event:event_table:by_time}. Each row in this table is one \emph{event record}. Each column is an \emph{attribute} where the column header defines the \emph{attribute name}. The contents of a table cell for event $e$ in column $a$ is the \emph{attribute value} event $e$ has for attribute $a$.

The event table can be considered as ``raw'' data as besides providing attributes per event, the data has no further structure. Specifically notice that no traces are recognizable in this event table.

\begin{table}\centering
\includegraphics[width=\linewidth]{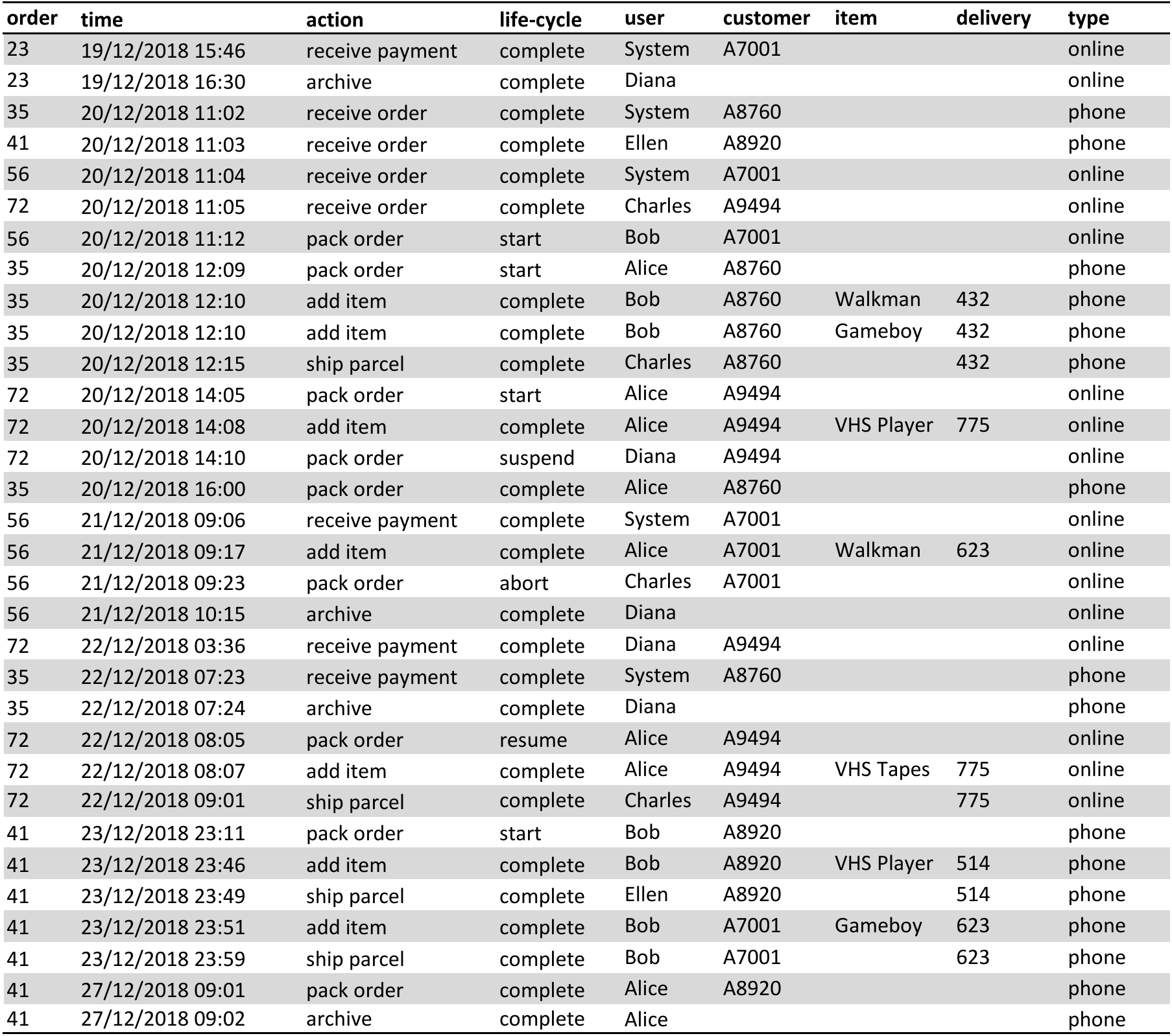}
\caption{Event Table}\label{tab:event:event_table:by_time}
\end{table}

The following definitions formally define events described by attributes and an event table.
\begin{itemize}
\item Let $\mathit{AN}$ be a set of \emph{attribute names}.
\item Let $\mathit{Val}$ a set of \emph{values}.
\item Let $\mathcal{E}$ be the universe of events.
\end{itemize}

\begin{definition}[Event]\label{def:event}
An \emph{event} $e \in \mathcal{E}$ describes that a specific discrete observation has been made (by a sensor, a system, a human observer, etc.). The observation itself is described by attribute-value pairs through the partial\footnote{A partial function does not have a value for each argument} function $\pi: \mathcal{E} \times \mathit{AN} \nrightarrow \mathit{Val}$.

For each event $e \in  \mathcal{E}$ and each attribute name $a \in \mathit{AN}$, $\pi(e,a) = v$ defines the value $v$ of attribute $a$. We write $\pi(e,a) = \perp$ if attribute $a$ is undefined for $e$ (has no value). We also write $\pi_a(e) = v$ or $e.a = v$ for $\pi(e,a) = v$.

For each event $e \in  \mathcal{E}$, we require that
\begin{itemize}
\item the attribute $\mathit{time}$ is defined, i.e., $\pi_\mathit{time}(e) \neq \perp$.
\item $e$ carries a value $\pi_a(e) \neq \perp$ for some other attribute $a \in \mathit{AN}, a \neq \mathit{time}$.
\end{itemize}
\end{definition}
In other process mining literature, you also find the notation $\#_a(e) = v$ instead of $\pi_a(e) = v$ or $e.a = v$ to describe that event $e$ has attribute $a$ with value $v$.

By the two requirements on each event $e \in  \mathcal{E}$ in Def.~\ref{def:event}, we ensure that each event has a timestamp $\pi_\mathit{time}(e)$ and records at least \emph{one} meaningful observation $\pi_a(e)$ (but it can record more). To be able to analyze processes in a meaningful way, we need the events we analyze to share some common ground: They should refer to the same kinds of observations, i.e., share some attributes. Therefore, an event table is a sequence of events, that all have the same attribute $a$ defined. We can think of attribute $a$ as the activity name or measurement that was recorded.
\begin{definition}[Event Table]\label{def:event_table}
An \emph{event table} is a finite sequence $\mathit{ET} = \trace{ e_1,\ldots,e_n}$ of events $e_1,\ldots,e_n \in \mathcal{E}$ of events with $\pi_a(e_i) \neq \perp$ for some attribute $a \in \mathit{ET}$ and all $1 \leq i \leq n$.

We write $e_i \in \mathit{ET}, 1 \leq i \leq n$ when referring to an event in $\mathit{ET}$.
\end{definition}

Definitions~\ref{def:event} and \ref{def:event_table} define the absolute bare minimum for analyzing events: all events $e$ have a timestamp $\pi_\mathit{time}(e)$ and record the some observation (or value) $\pi_a(e_i)$. In this bare form, an event table could even specify a time-series. However, most events carry many additional attributes which we exploit in process mining.

Table~\ref{tab:event:event_table:by_time} shows an event table according to Def.~\ref{def:event} and Def.~\ref{def:event_table}.

\begin{exercise}
Choose any event from Table~\ref{tab:event:event_table:by_time} and give its formal definition according to Def.~\ref{def:event}.
\end{exercise}

Strictly speaking, Definition~\ref{def:event_table} does not define an event \emph{table} in the sense of the data model of relational databases, but rather just a finite stream of events of attribute-value pairs. However, the table format representation is convenient and data in this form is often stored and exchanged using the \emph{Comma Separated Value (CSV)} format.

\begin{exercise}
What are the differences between Definition~\ref{def:event_table} and the data model of relational databases?
\end{exercise}

We can reorder the events/rows in an event table to better understand its contents. Table~\ref{tab:event:event_table:by_orderid} reorders the events of Table~\ref{tab:event:event_table:by_time} by grouping them by attribute \emph{order} and the sorting all events per order on attribute \emph{time}. In Table~\ref{tab:event:event_table:by_orderid}, we added a column assigning each event a unique identifier to be able to refer to them individually, e.g., $e_1$ is the first event in this table.

In this sorted event table, we can start recognizing the traces we discussed in Sect.~\ref{sec:event:event_data}. However, the traces are no objects yet in their own right. We discuss how to obtain traces and structured event logs next.

\section{Extracting Structured Event Logs from an Event Table}\label{sec:event:structured_log}

An event table only records for each event its timestamp and some observation such as an activity name. The essential difference between an event table and an event \emph{log} is the presence of an additional attribute called the \emph{case identifer}. It allows to group events into cases and traces and compare multiple sequences of events to each other.

\subsection{Entities and Case}

We use the term \emph{Case} to refer to an \emph{entity} or \emph{object} that we are ``tracking'' over time in terms of the events in which this entity is involved.

For example, for the first event in Table~\ref{tab:event:event_table:by_orderid} we can recognize that three \emph{types of entities} were involved.
\begin{itemize}
\item \emph{order} (for which we find the order id ``23'' as attribute value)
\item \emph{user} (for which we find the user name ``System'' as attribute value)
\item \emph{customer} (for which we find an identifer ``A7001'' as attribute value)
\end{itemize}

Other events also refer to a fourth entity type \emph{delivery}.

In contrast, the attribute \emph{item} does not refer to an entity type because its values describe sets or classes of similar objects but do not identify a unique entity or object. Recognizing which attributes of an event refer to entity types requires domain knowledge or additional context information.

To obtain a structured event log from an event table, we have to recognize from all attribute names the entity types, and then select one these attributes $c$ referring to an entity type as the \emph{case identifier} attribute. The attribute values for $c$ we find among all events are the cases we find in the data.
\begin{definition}[Case identifier, cases]\label{def:case_id}\label{def:case}
Let $\mathit{ET} = \trace{e_1,\ldots,e_n}$ be an event table.

The set of attribute names in $\mathit{ET}$ is
$$\mathit{AN}(ET) = \{ a \in \mathit{AN} \mid \exists e_i \in \mathit{ET}, \pi_a(e_i) \neq \perp \}.$$

If we select an attribute $\mathit{id} \in \mathit{AN}(ET)$ as \emph{case identifier}, then $$\mathit{Cases}(\mathit{ET},\mathit{id}) = \{ \pi_{\mathit{id}}(e_i) \mid e_i \in \mathit{ET}\}$$ is the set of cases for this case identifier.
\end{definition}
By choosing one entity type as case identifier, we decide to reformat the event data in a way that ``tracks'' what has happened to all entities of this type.

Note that Definition~\ref{def:case_id} allows to pick any attribute as case identifier, not just those that refer to entity types. For example, we could even pick attribute name $\mathit{action} \in \mathit{AN}(ET)$. The next steps in building a structured event log work with any chosen case identifier. However, the subsequent analysis entirely depends on how sensible this choice of a case identifier was for the particular analysis question. In other words, we have to understand which analysis question we try to answer, and then identify the corresponding attribute name (e.g., of an entity type of interest) that we want to use as case identifier.

This also means that at this point we implicitly require that each event $e$ has \emph{three} mandatory attributes that are different from each other (i.e., we do not choose $a = c$ or $a = c = \mathit{time})$:
\begin{enumerate}
  \item the \emph{timestamp} $\pi_{\mathit{time}}(e)$ (see Def.~\ref{def:event})
  \item a recorded action or \emph{activity} $\pi_{\mathit{a}}(e)$ (see Def.~\ref{def:event_table})
  \item a \emph{case identifier} $\pi_{\mathit{c}}(e)$ (see Def.~\ref{def:case_id})
\end{enumerate}
However, except for $\pi_{\mathit{time}}(e)$, activity and case identifier are \emph{not} pre-determined by the event table. They are choices we make.

\begin{exercise}
Can two events happen at the same time? Do they have to be in different cases? Do they have to have different activities?
\end{exercise}

\begin{table}\centering
\includegraphics[width=\linewidth]{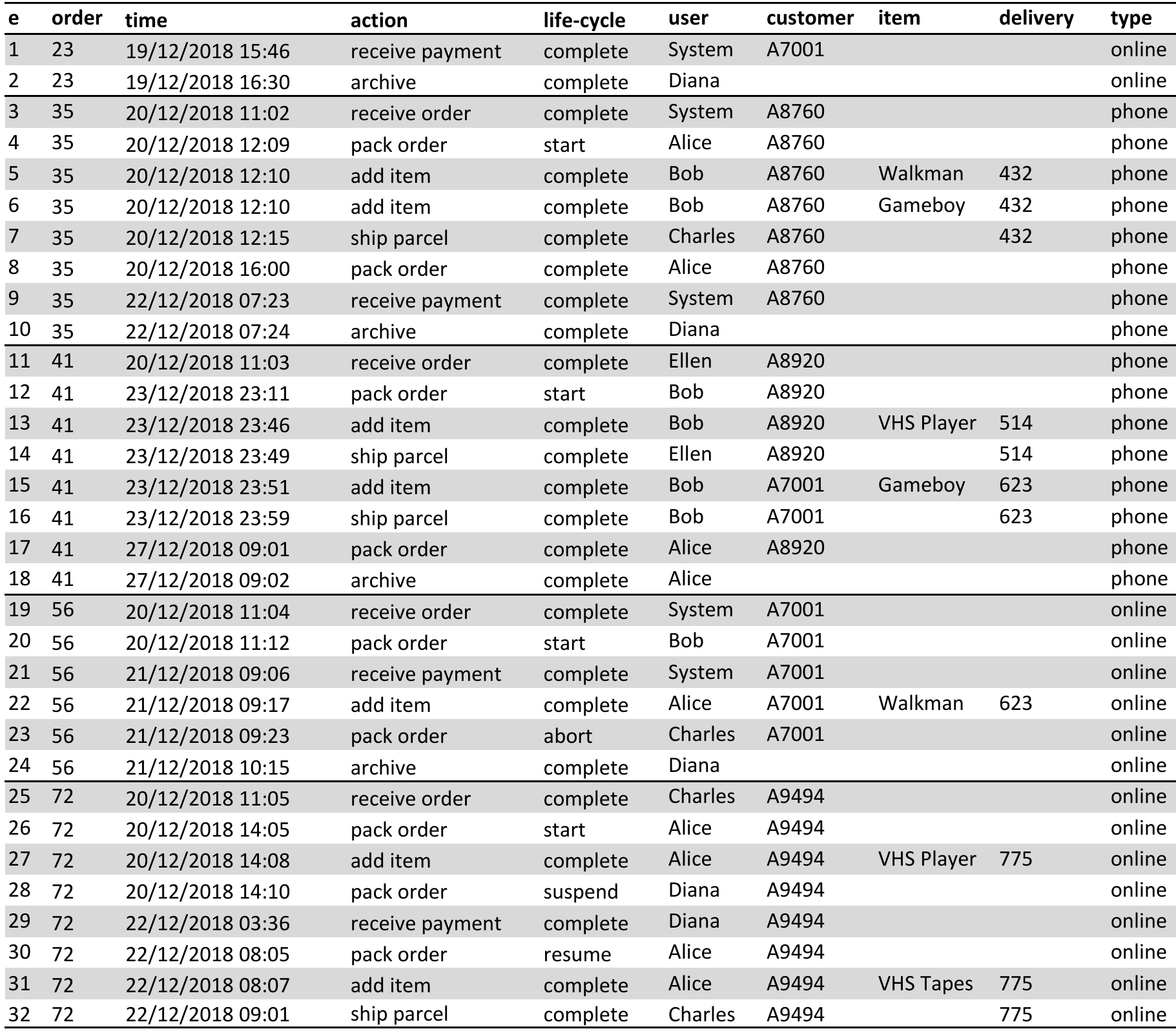}
\caption{Event Table}\label{tab:event:event_table:by_orderid}
\end{table}

For the example of Table~\ref{tab:event:event_table:by_orderid}, we can see four candidates for case identifiers based on the entity-types we found in the event table: $\mathit{order}$, $\mathit{delivery}$, $\mathit{user}$, and $\mathit{customer}$.

\begin{exercise}
What are the cases for case identifier $\mathit{delivery}$ in Table~\ref{tab:event:event_table:by_orderid}?
\end{exercise}

If we have selected an attribute $id$ as case identifier, then we say an event $e$ is \emph{correlated} to a case $c$ if its $id$-attribute refers to $c$, i.e., $\pi_{id}(e) = c$.
\begin{definition}[Correlation to a case]\label{def:correlation}
Let $\mathit{ET}$ be an event table. Let $c \in \mathit{Cases}(\mathit{ET},\mathit{id})$ be a case for a case identifier $\mathit{id} \in \mathit{AN}(ET)$.

Event $e \in \mathit{ET}$ is \emph{correlated} to $c$ iff $\pi_{id}(e) = c$.
The set of all events correlated to $c$ is $$\mathit{corr}(\mathit{ET},id,c) = \{ e \in \mathit{ET} \mid \pi_{id}(e) = c\}.$$
\end{definition}
For example, events $e_1$ and $e_2$ are correlated to $order=23$ in Table~\ref{tab:event:event_table:by_orderid}. Note that if an event $e$ does not have attribute $id$ defined, i.e., $\pi_{id}(e) = \perp$, then it is not correlated to any case of this case identifier.

\begin{exercise}
Which events are correlated to $order=35$?
\end{exercise}

\begin{exercise}
Which events are correlated to $delivery=623$?
\end{exercise}

\begin{exercise}
What happens when we choose $\mathit{time}$ to be an activity name or a case identifier?
\end{exercise}

If all events correlated to a case $c$ carry the \emph{same} value $v$ for an attribute $x$, then we call $x$ a \emph{case attribute} of $c$.
\begin{definition}[Case attribute]\label{def:case_attribute}
Let $\mathit{ET}$ be an event table. Let $c \in \mathit{Cases}(\mathit{ET},\mathit{id})$ be a case for a case identifier $\mathit{id} \in \mathit{AN}(ET)$.

Attribute $x \in \mathit{AN}(ET)$ is a \emph{case attribute} of $c$ iff for all $e,e' \in \mathit{corr}(\mathit{ET},id,c)$ holds $\pi_x(e) = \pi_x(e') = v \neq \perp$. We then lift the function $\pi(.)$ from events to cases and write $\pi_x(c) = v$.

Attribute $x$ is a \emph{global} case attribute iff it is a case attribute for every case $c \in \mathit{Cases}(\mathit{ET},\mathit{id})$.
\end{definition}
The case identifier is \emph{always} a global case attribute. A global case attribute has to be defined for each case, but each case can have its own value.

\begin{exercise}
What are the case attributes of Table~\ref{tab:event:event_table:by_orderid} for case identifier $order$?
\end{exercise}

\subsection{Trace}

A \emph{trace} is the sequence of events correlated to a case and ordered by time. For example, the trace of $order=23$ in Table~\ref{tab:event:event_table:by_orderid} is $\trace{e_1,e_2}$.

\begin{definition}[Trace of a case]\label{def:trace}
Let $\mathit{ET} = \trace{e_1,\ldots,e_n}$ be an event table. Let $\mathit{id} \in \mathit{AN}(ET)$ be the selected case identifier.

A sequence $\trace{e_1,\ldots,e_k}$ of events is a \emph{trace} of case $c \in \mathit{Cases}(\mathit{ET},\mathit{id})$ iff
\begin{enumerate}
  \item $\{e_1,\ldots,e_k\} = \mathit{corr}(\mathit{ET},id,c)$, i.e., it consists of all events of $\mathit{ET}$ correlated to $c$, and
  \item for each $i = 1,\ldots,k-1$ holds $\pi_{\mathit{time}}(e_i) \leq \pi_{\mathit{time}}(e_{i+1})$, i.e., events are ordered by time.
\end{enumerate}
\end{definition}
Note that there may be more than one way to sequentialize the events $\{e_1,\ldots,e_k\} = \mathit{corr}(\mathit{ET},id,c)$ correlated to a case $c$. This happens where two or more events $e_i,e_{i+1}$ have the same time-stamp $\pi_{\mathit{time}}(e_i) = \pi_{\mathit{time}}(e_{i+1})$.

\begin{exercise}
What is the trace of $order=35$?
\end{exercise}

\begin{exercise}
What is the trace of $delivery=623$?
\end{exercise}

\subsection{Structured Event Log}

A structured event log is a set of cases where each case is associated with exactly one trace for this case as a case attribute.
\begin{definition}[Structured Event Log]\label{def:structured_event_log}
Let $\mathit{ET} = \trace{e_1,\ldots,e_n}$ be an event table. Let $\mathit{id} \in \mathit{AN}(ET)$ be the selected case identifier.

The \emph{structured} event log $L$ is the set $L = \mathit{Cases}(\mathit{ET},\mathit{id})$ of cases for case identifier $\mathit{id}$ so that additionally each case $c \in L$ gets assigned a trace $\trace{e_1,\ldots,e_k}$ of $c$ as trace attribute $\pi_{\mathit{trace}}(c) = \trace{e_1,\ldots,e_k}$.
\end{definition}
For example, the structured event log of Table~\ref{tab:event:event_table:by_orderid} has the cases $L = \{23,35,41,56,72\}$ for $order$ and the following traces:
\begin{itemize}
  \item $\pi_{trace}(23) = \trace{ e_1,e_2 }$ where
  \begin{itemize}
  \item $e_1$ has
      \begin{itemize}
      \item $\pi_{order}(e_1) = 23$
      \item $\pi_{time}(e_1) = \text{19/12/2018 15:46}$
      \item $\pi_{action}(e_1) = \text{receive payment}$
      \item $\ldots$
      \end{itemize}
  \item $e_2$ has
    \begin{itemize}
      \item $\pi_{order}(e_2) = 23$
      \item $\pi_{time}(e_2) = \text{19/12/2018 16:30}$
      \item $\pi_{action}(e_2) = \text{archive}$
      \item $\ldots$
      \end{itemize}
  \end{itemize}
  \item $\pi_{trace}(35) = \trace{ e_3,e_4,e_5,\ldots,e_{10}}$
  \item $\pi_{trace}(41) = \trace{ e_{11},e_{12},e_{13},\ldots,e_{18}}$
  \item $\pi_{trace}(56) = \trace{ e_{19},\ldots,e_{24}}$
  \item $\pi_{trace}(72) = \trace{ e_{25},\ldots,e_{32}}$
\end{itemize}

A structured event log has a simple hierarchical structure. At the top-level are the cases $L = \{c_1,\ldots,c_k\} = \mathit{Cases}(\mathit{ET},\mathit{id})$. Each case has case attributes as ``children'', one of them is the trace $\pi_{\mathit{trace}}(c)$. Each event $e$ in a trace has event attributes as children, including $\pi_{time}(e)$ (time-stamp), $\pi_{a}(e)$ (the observed activity), and $\pi_{c}(e)$ (the case identifier).
\begin{enumerate}
\item A structured event log $L$ consists of a set of cases $L = \{c_1,\ldots,c_n\} \subseteq \mathit{Val}$, i.e., values for some case identifier.
\item Each case $c \in L$ defines a trace $\pi_{\mathit{trace}}(c) = \trace{e_1,\ldots,e_k} \in \mathcal{E}^*$ as a sequence of events ordered by time, i.e., $\pi_{\mathit{time}}(e_i) \leq \pi_{\mathit{time}}(e_{i+1})$ for each $i = 1,\ldots,k-1$.
\item The events in $\pi_{\mathit{trace}}(c)$ are all correlated to the  case, i.e., $\pi_{\mathit{id}}(e_i) = c$. However, most XES event logs do not store the case identifier as an event attribute again.
\item There is at least one attribute $a$ (e.g., the activity name) defined by each event $\pi_a(e)$ in each trace $e \in \pi_{\mathit{trace}}(c), c\in L$.
\item Cases do not share events, i.e., there is no event $e \in  \mathcal{E}$\\ with $e \in \pi_{\mathit{trace}}(c), \pi_{\mathit{trace}}(c'), c,c' \in L, c \neq c'$.
\end{enumerate}
This hierarchical structure is formalized in the XES-standard~\cite{Acampora2017xes,XES16,xes10}.
See also other formalizations of event logs~\cite{DBLP:books/sp/Aalst16}

\begin{exercise}
Provide the cases and traces of the structured event log of Table~\ref{tab:event:event_table:by_orderid} for case identifier \emph{delivery}.
\end{exercise}

\begin{exercise}
Which other meaningful structured event logs can you extract from Table~\ref{tab:event:event_table:by_orderid}?
\end{exercise}

\section{Event Classifiers and Simple Event Logs}\label{sec:event:classifiers}

The nested hierarchy of a structured event log contains all information about all events. However, analysis techniques operating on events, prefer a flat data structure where
\begin{itemize}
\item a structured event $e \in \mathcal{E}$ with its various attributes is represented by a single attribute value $\pi_{a}(e)$, e.g., the activity name,
\item a structured case $c$ with its various case attributes is not represented its trace $\pi_{trace}(c) = \trace{e_1,\ldots,e_k}$ but rather in its simplified form $\trace{\pi_{a}(e_1),\ldots,\pi_{a}(e_k)}$.
\end{itemize}
For example, we can transform $\trace{e_1,e_2}$ of Table~\ref{tab:event:event_table:by_orderid} into $\trace{ \text{receive payment}, \text{archive}}$. This representation allows easily searching for patterns in the sequences of activity names.

\subsection{Event Classifiers and Event Classes}

However, as for Definition~\ref{def:case_id} of the case identifier, events do not have a canonical or standard attribute $a$ by which it \emph{must} be represented in this simplified way. Rather, we again can pick.

Literature introduces for this purpose the definition of an \emph{event classifier}.

\begin{definition}[Event Classifier] \label{def:event_classifier}\label{def:event_class}
An event classifier is a function with signature $$\mathit{class} : \mathcal{E} \to \mathit{Value}$$ that maps each event to a value. The value $\mathit{class}(e)$ is called the \emph{event class}. Any two events $e,e'$ with $\mathit{class}(e) = \mathit{class}(e')$ belong to the same event class, which means they describe the same kind of observation.
\end{definition}

Usually, the event classifier is defined over event attributes which can be a single attribute or a combination of attributes. For example,
\begin{itemize}
\item The standard event classifier is the \emph{activity name classifier} $\mathit{class}_{act}(e) = \pi_{a}(e)$ where $\mathit{a} \in \mathit{AN}$ is the attribute we identify as the activity name.

    For Table~\ref{tab:event:event_table:by_orderid}, the \emph{activity name classifier} is  $\mathit{class}_{act}(e) = \pi_{action}(e)$. For example, events $e_4, e_8$ have the same activity name class $\mathit{class}_{act}(e_4) = \mathit{class}_{act}(e_8) = \text{pack order}$.

\item The \emph{activity+lifecycle} classifier combines the activity name $a$ with the event life-cycle attribute $lc$ (if it exists in the event log), i.e., $\mathit{class}_{act+lifecycle}(e) = (\pi_{a}(e),\pi_{lc}(e))$.

    For Table~\ref{tab:event:event_table:by_orderid}, the \emph{activity+lifecycle} classifier is\\  $\mathit{class}_{act+lifecycle}(e) = (\pi_{action}(e),\pi_{\text{life-cycle}}(e))$.

    For example, events $e_4$ belong to different event classes for this classifier: $\mathit{class}_{act+lifecycle}(e_4) = (\text{pack order},\text{start})$ and\\
    $\mathit{class}_{act+lifecycle}(e_8) = (\text{pack order},\text{complete})$.

    Event classes over multiple attributes are also represented with a `+', e.g., $\mathit{class}_{act+lifecycle}(e_4) = \text{pack order+start}$.

\item The \emph{resource} classifier $\mathit{class}_{res}(e) = \pi_{r}(e)$ where $\mathit{r} \in \mathit{AN}$ is the attribute deferring to the user, machines, or resource that participated in the event.

    For Table~\ref{tab:event:event_table:by_orderid}, the \emph{resource} classifier is  $\mathit{class}_{res}(e) = \pi_{user}(e)$.
    For example, events $e_4$ and $e_8$ belong to the same event resource event class: $\mathit{class}_{res}(e_4) = \mathit{class}_{res}(e_8) = \text{Alice}$.
\end{itemize}
We can in principle choose any combination of attributes for the event classifier. This essentially corresponds to \emph{feature selection} in data mining: we choose the event attributes we think are most relevant for the analysis task at hand. If the event has no value defined for the selected event classifier, e.g., $\mathit{class}_{item}(e) = \pi_{item}(e)$ and $\mathit{class}_{item}(e_1) = \perp$, then the event will be omitted from the analysis.

It is also possible to derive new event attributes based on other events in the trace or even the entire event log, and to use these subsequently as event classifiers. This would correspond to \emph{feature engineering}.

\begin{exercise}
Identify another meaningful event classifier from Table~\ref{tab:event:event_table:by_orderid}.
\end{exercise}

\begin{definition}[Event Classes of an Event Log]\label{def:event_classes}
Given a structured event log $L$ (according to Def.~\ref{def:structured_event_log}) and an event classifier $\mathit{class}$, the \emph{set of event classes} in $L$ is the set $\Sigma_\mathit{class}(L) = \{ \mathit{class}(e) \mid c \in L, e \in \pi_{trace}(c), \mathit{class}(e) \neq \perp \}$.
\end{definition}

\begin{exercise}
What are the event classes of the event log in Table~\ref{tab:event:event_table:by_orderid} for the classifier $\mathit{class}(e) = \pi_{\text{customer}}(e)?$
\end{exercise}

\subsection{Simple Event Log}

If we have fixed an event classifier $\mathit{class}$, we can represent each trace in a log $L$ by the sequence of event classes, e.g., the sequence of activity names. However, we omit all $\perp$ values.
\begin{definition}[Simple Trace]
Let $L$ be a structured event log, let $\trace{e_1,\ldots,e_k} = \pi_{trace}(c),c\in L$ be a trace. Let $\mathit{class}$ be an event classifier.

The \emph{simple trace} of $c$ is the sequence $$\mathit{simple}_\mathit{class}(c) = \trace{\mathit{class}(e_1),\ldots,\mathit{class}(e_k)}|_{\Sigma_\mathit{class}(L)}$$
where we replace each event $e_i$ by $\mathit{class}(e_i)$ and then project\footnote{We write $\sigma|_{\Sigma'}$ for the projection of a trace $\sigma \subseteq \Sigma^*$ onto a subset $\Sigma' \subseteq \Sigma$ of some alphabet.} the resulting sequence onto all valid event classes $\Sigma_\mathit{class}(L)$, i.e., all values that are not $\perp$.
\end{definition}

The simple trace for case $23$ and the activity event classifier $\mathit{class}_{act}(e) = \pi_{action}(e)$ is

$$\trace{\text{receive payment},\text{archive}}.$$

The simple trace for case $23$ and the activity event classifier $\mathit{class}(e) = \pi_{customer}(e)$ is

$$\trace{\text{A7001}}.$$

We obtain the simple event log of $L$ by collecting all simple traces of all cases $L$ in a multiset. Recall from Sect.~\ref{sec:event:event_data} that $\sigma \in \Sigma^*$ is a finite sequence of activity names and $\bag(\Sigma^*)$ is a multi-set (bag) of finite sequences.
\begin{definition}[Simple Event Log]
Let $L$ be a structured event log. Let $\mathit{class}$ be an event classifier. Let $\Sigma = \Sigma_\mathit{class}(L)$

The simple event log is the multiset $\mathit{simple}_\mathit{class}(L) = L' \in \bag(\Sigma^*)$ where $L'(\sigma) = | \{ c \in L \mid \mathit{simple}_\mathit{class}(c) = \sigma \} |$, i.e., there are as many copies of $\sigma$ as there are cases which have the same simple trace $\mathit{simple}_\mathit{class}(c) = \sigma$.
\end{definition}
The simple event log of Table~\ref{tab:event:event_table:by_orderid} for the \emph{action} classifier is (we abbreviate each action name for succinctness):
\begin{displaymath}
\begin{array}{rl}
L' = [ & \trace{ \text{RP},\text{AR} }^1,\\
       & \trace{ \text{RO},\text{PO},\text{AI},\text{AI},\text{SP},\text{PO},\text{RP},\text{AR} }^1,\\
       & \trace{ \text{RO},\text{PO},\text{AI},\text{SP},\text{AI},\text{SP},\text{PO},\text{AR} }^1,\\
       & \trace{ \text{RO},\text{PO},\text{RP},\text{AI},\text{PO},\text{AR} }^1,\\
       & \trace{ \text{RO},\text{PO},\text{AI},\text{PO},\text{RP},\text{PO},\text{AI},\text{SP}  }^1].
\end{array}
\end{displaymath}
The simple traces in a simple event log are also called \emph{trace variants} of the event log as they show the principle ways the object that is tracked by the case identifier ``moves'' through the data.

\begin{exercise}
What is the simple event log for the \emph{activity+life-cycle} classifier?
\end{exercise}
\begin{exercise}
What is the simple event log for the \emph{item} classifier?
\end{exercise}

Note that each simple trace is a finite sequence $\sigma \in \Sigma^*$ over some alphabet of event classes $\Sigma = \Sigma_\mathit{class}(L)$ and that the simple event log is a multiset of simple traces. We now have a complete procedure for obtaining a simple event log, as outlined in Sect.~\ref{sec:event:event_data} from event data as it is recorded in practice, i.e., an event table.
\begin{enumerate}
\item Find meaningful entity identifiers in the attributes of the event table that correspond to your analysis question.
\item Select one entity identifier as case identifier $\mathit{id}$.
\item Construct the structured event log $L$ for this case identifier by correlating events and ordering them over time.
\item Find meaningful event attributes to summarize or classify the observation that is recorded in the event.
\item Select or define one event classifier $\mathit{class}$.
\item Derive the simple event log from $L$ for this event classifier $\mathit{class}$.
\end{enumerate}
Given an event table $\mathit{ET}$, any simple event log is fully defined by two decision: the case identifier $\mathit{id}$ and the event classifier $\mathit{class}$. However, these two choices are powerful and allow you to derive many different views.
\begin{exercise}
What is the simple event log for case identifier \emph{customer} and event classifier \emph{order}?
\end{exercise}

\begin{figure}\centering
\includegraphics[width=\linewidth]{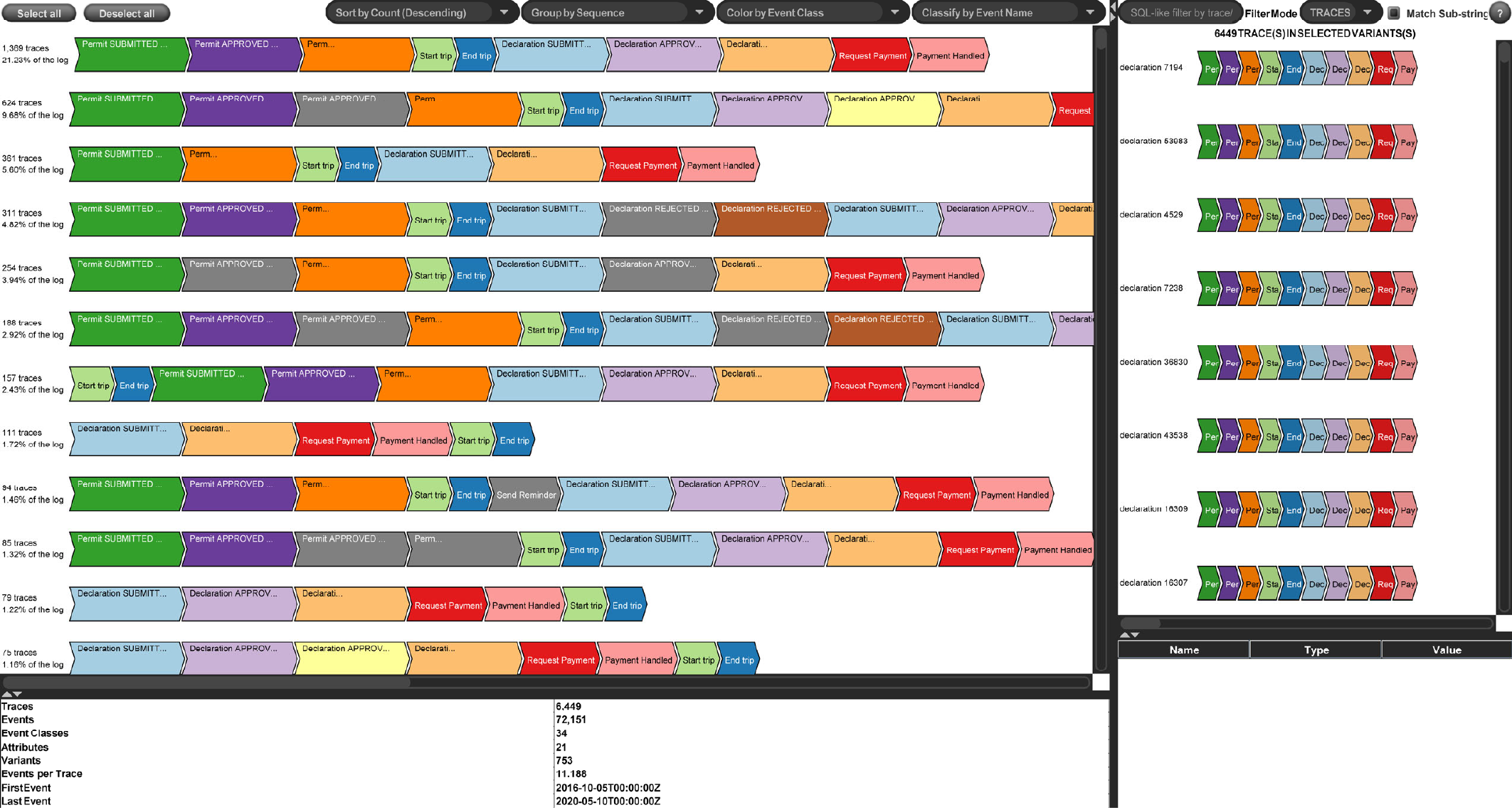}
\caption{The Explore Event Log visualizer of ProM}\label{fig:event:prom_explore_event_log}
\end{figure}

Almost all process mining software contains a view to visualize the event log in the form of a simple event logs; for example the ``Explore Event Log'' visualizer of ProM\footnote{http://www.promtools.org/} shown in Fig.~\ref{fig:event:prom_explore_event_log} visualizes event logs as simple event logs and allows identifying patterns through color-coding the event classes.

\section{Pre-Processing Event Logs}\label{sec:event:pre-processing}

Analyzing event data, just like any data analysis, requires pre-processing to remove data points that are not relevant for the specific analysis question at hand.

\begin{figure}
  \centering
  \includegraphics[width=\linewidth]{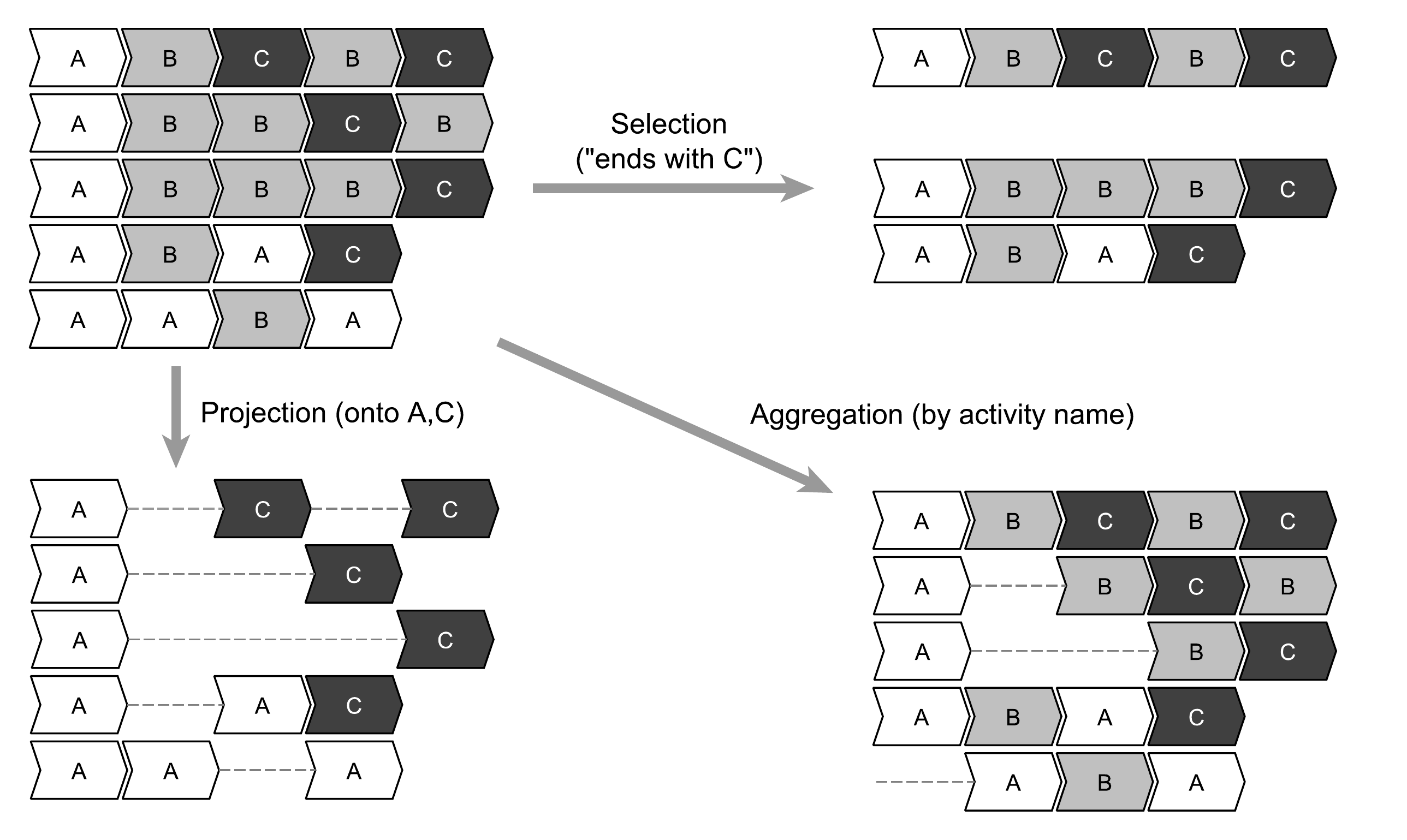}
  \caption{Pre-Processing Operations on Event Logs}\label{fig:event:pre-processing}
\end{figure}

There are three basic pre-processing operations on event logs that allow us to reduce or ``filter'' the data in three fundamentally different ways. Most other pre-processing operations are a combination of these three operations. They are defined on the data model of the structured event log (Def.~\ref{def:structured_event_log}).
\begin{enumerate}
\item \emph{Selection} of traces reduces the set of cases in $L$ to those that satisfy a specific property. All other cases are removed. The pre-processed log $L'$ contains just a subset of the cases in $L$, i.e., $L' \subseteq L$ and each cases keeps all its properties, especially all events in its trace.

    Figure~\ref{fig:event:pre-processing} (top) illustrates the selection of $L$ to all cases whose traces end with an event with activity name $C$. The resulting log $L'$ does not contain the cases whose traces end with $B$ or $A$.
\item \emph{Projection} removes from each trace in $L$ all events that do \emph{not} satisfy a particular property. The resulting event log $L'$ keeps all its cases, but their traces may contain fewer events or even be empty.

    Figure~\ref{fig:event:pre-processing} (left) illustrates the projection of $L$ to all events with activity attribute $A$ or $C$. The resulting event log does not contain any event with activity attribute $B$ anymore.
\item \emph{Aggregation} groups in each trace multiple subsequent events $e_1,\ldots,e_k$ with the same property into a new event $e^*$ whose properties are derived from $e_1,\ldots,e_k$; $e_1,\ldots,e_k$ are then replaced by $e^*$. The pre-processed log $L'$ keeps all its cases, but the traces may have fewer events and may contain a new aggregated event with new properties that were not explicitly visible in $L$.

    Figure~\ref{fig:event:pre-processing} (bottom right) illustrates the aggregation of subsequent events with the same activity name. For example, the subsequence $\trace{B,B}$ in the second case was replaced by a single $B$.
\end{enumerate}

\begin{definition}[Selection]\label{def:preprocessing:selection}
Let $L$ be a structured event log. Let $\varphi(c)$ be a predicate over the case attributes and event attributes of $L$. The \emph{selection} of $L$ wrt. $\varphi$ is the subset $$\mathit{Select}_\varphi(L) = \{ c \in L \mid \varphi(c) = true \}.$$
\end{definition}
Here are several example selection predicates for the event log in Table~\ref{tab:event:event_table:by_orderid}:
\begin{itemize}
  \item $\varphi_1(c) \equiv \pi_{type}(c) = \text{online}$ (only cases of type ``online'')
  \item $\varphi_2(c) \equiv \pi_{trace}(c) = \trace{e_1,\ldots,e_n} \logAnd \pi_{action}(e_1) = \text{receive order}$ (only cases starting with ``receive order'')
  \item $\varphi_3(c) \equiv \pi_{trace}(c) = \trace{e_1,\ldots,e_n} \logAnd \pi_{time}(e_n) - \pi_{time}(e_1) < 24h$ (only cases completing within 24 hours)
  \item $\varphi_4(c) \equiv |\{ c'\in L \mid \mathit{simple}_\mathit{class}(c) = \mathit{simple}_\mathit{class}(c')\}| \geq 10$ for some event classifier $\mathit{class}$ (only cases whose trace variant, i.e., simple trace, occurs at least 10 times in the event log)
\end{itemize}
Note that $\varphi_4(c)$ is not purely local to the case $c$ but rather ``reaches out'' into the entire event log $L$.

\begin{exercise}
Which cases are selected by $\varphi_1(c)$-$\varphi_4(c)$?
\end{exercise}

\begin{definition}[Projection]\label{def:preprocessing:projection}
Let $L$ be a structured event log. Let $\psi(e)$ be a predicate over the event attributes of $L$.

Let $\sigma = \trace{e_1,\ldots,e_n} = \pi_{trace}(c), c\in L$ be a trace. The \emph{projection} of $\sigma$ onto $\psi$ is the projection of $\sigma$ onto all events $e_i$ where $\psi(e_i) = true$, i.e.,  $$\mathit{Proj}_\psi(\sigma) = \trace{e_1,\ldots,e_n}|_{\psi(e_i) = true}.$$

We obtain the \emph{projection} of $L$ into $\psi$, written  $\mathit{Proj}_\psi(L)$, by setting $\pi_{trace}(c) \df \mathit{Proj}_\psi(\pi_{trace}(c))$.
\end{definition}
Here are several example projection predicates for the event log in Table~\ref{tab:event:event_table:by_orderid}:
\begin{itemize}
  \item $\psi_1(e) \equiv \pi_{\text{life-cycle}}(e) = \text{complete}$ (only ``complete'' events)
  \item $\psi_2(e) \equiv \pi_{\text{delivery}}(e) \neq \perp$ (only events with a reference to a delivery)
  \item $\psi_3(e) \equiv \pi_{\text{type}}(e) = \text{online}$
  \item $\psi_4(e) \equiv \pi_{\text{user}}(e) \in \{\text{Alice}, \text{Bob}\}$ (only events where Alice or Bob are involved)
  \item $\psi_5(e) \equiv c = \pi_{order}(e) \logAnd
  \pi_{trace}(c) = \trace{e_1,\ldots,e_n} \logAnd e = e_i \logAnd \forall j = i+1,\ldots,n \logAnd \pi_{action}(e_i) \neq \pi_{action}(e_j)$ (only the last occurrence of each activity in a trace)
  \item $\psi_6(e) \equiv |\{ e' \mid c' \in L, e' \in \pi_{trace}{c'}, \pi_{act}(e) = \pi_{act}(e')| \geq 5$ (only events of activities which occur at least 5 times in the event log $L$).
\end{itemize}
Note that when we constructed the event log from the event table, each event $e$ had the chosen case identifier $id$ as event attribute, i.e., $\pi_{id}(e) = c$ refers to the case. When constructing the event log, we used the value $c$ to construct the case itself. This means, we can ``reach'' the case $c$ from an event $e$, and once we have the case $c$, we can ``reach'' the entire trace $\pi_{trace}(c)$ that also contains $e$. We use this in $\psi_5(e)$ to reason about whether $e$ is not the last event in the trace of the same activity.

Such projection attributes are not possible in all process mining software. It can only be defined if the event $e$ actually has a reference to the case $c$ and the data structure in which the event is stored allows to resolve this reference. Similarly, $\psi_6(e)$ requires that the entire event log (or statistics about the event log) are accessible.

\begin{exercise}
Apply $\mathit{Proj}_{\psi_2}(L)$ on the event log of Table~\ref{tab:event:event_table:by_orderid} and derive the simple event log for the activity name event classifier.
\end{exercise}

\begin{exercise}
What is the difference between $\mathit{Select}_\mathit{selectP_1}(L)$ and $\mathit{Proj}_\mathit{projP_3}(L)$?
\end{exercise}

For aggregation, we do not provide a full formal definition, but outline what has to be defined. Aggregation in a case $\mathit{Agg}_{g,r}(c)$ requires two functions $g$ and $r$:
\begin{itemize}
  \item A \emph{grouping} classifier $g : \mathcal{E} \to \mathit{Val}$ which maps each event to a value, similar to an event classifier.
  \item With $g$, we partition the trace $\pi_{trace}(c) = \sigma$ into maximal subsequences $\sigma_i \trace{e_{i_1},\ldots,e_{i_n}}$ so that $g(e_j) = g(e_{j+1})$ for all $i_1 \leq j < i_n$, e.g., all subsequences with the same activity name. This results in a sequence of $k$ such sub-sequences of various lengths, i.e., $g(\sigma) = \trace{ \sigma_1,\ldots,\sigma_k }$, for instance, $\trace{ \trace{e_1,e_2}, \trace{e_3}, \trace{e_4,e_5}}$
  \item A \emph{replacement} function $r : \mathcal{E}^+ \to \mathcal{E}$ that replaces any non-empty subsequence $\trace{e_i,\ldots,e_{i+m}}$ by a new event $r(\trace{e_i,\ldots,e_{i+m}}) = e'$ and defines the event attributes for $e'$ based on the attribute values of $\trace{e_i,\ldots,e_{i+m}}$; $r$ specifically has to set the timestamp of $e'$ to be within $\pi_{time}(e_i) \leq \pi_{time}(e') \leq \pi_{time}(e_{i+m})$. For a singleton sub-sequence $\trace{e_i}$, the replacement function should just return the event $e_i$, i.e., the event remains unchanged.
  \item $\mathit{Agg}_{g,r}(c)$ apply $r$ to each sub-sequence $\trace{e_i,\ldots,e_{i+m}}$ obtained from $g$ which results in a new sequence of events, i.e.,  $\mathit{Agg}_{g,r}(c) = \trace{r(\sigma_1),\ldots,r(\sigma_k)}$ where $g(\sigma) = \trace{ \sigma_1,\ldots,\sigma_k}$ and $\sigma = \pi_{trace}(c)$. For example,\\ $\trace{ r(\trace{e_1,e_2}), r(\trace{e_3}), r(\trace{e_4,e_5})} = \trace{e_{12},e_3,e_{45}}$.
  \item Set $\pi_{trace}(c) \df \mathit{Agg}_{g,r}(c)$
\end{itemize}
For example, we can use $g(e) = \pi_{action}(e)$ to find all subsequences where the same activity occurs repeatedly. In Table~\ref{tab:event:event_table:by_orderid}, this would be only $\trace{e_5,e_6}$. We can then define a replacement function $r(\trace{e_1,\ldots,e_k})$ where the new event $e'$ gets $\pi_{x}(e') = \pi_{x}(e_k)$ for all attributes $x$ defined for $e_k$, i.e., we replace the sequence by the last event. We could also define $\pi_{item}(e') = \{ \pi_{item}(e_i) \mid 1 \leq i \leq k \}$ to collect the items that were involved in these events into a set.

\begin{exercise}
What is the difference between aggregation where any subsequence is replaced by the last event and projection with $\psi_5(e)$?
\end{exercise}

All three event log pre-processing operations selection, projection, and aggregation always result in single event log. This allows to apply them in arbitrary combinations. For example, first project onto all events where ``Bob'' is involved and then aggregate on $\pi_{action}(e)$.

\begin{figure}
  \centering
  \includegraphics[width=\linewidth]{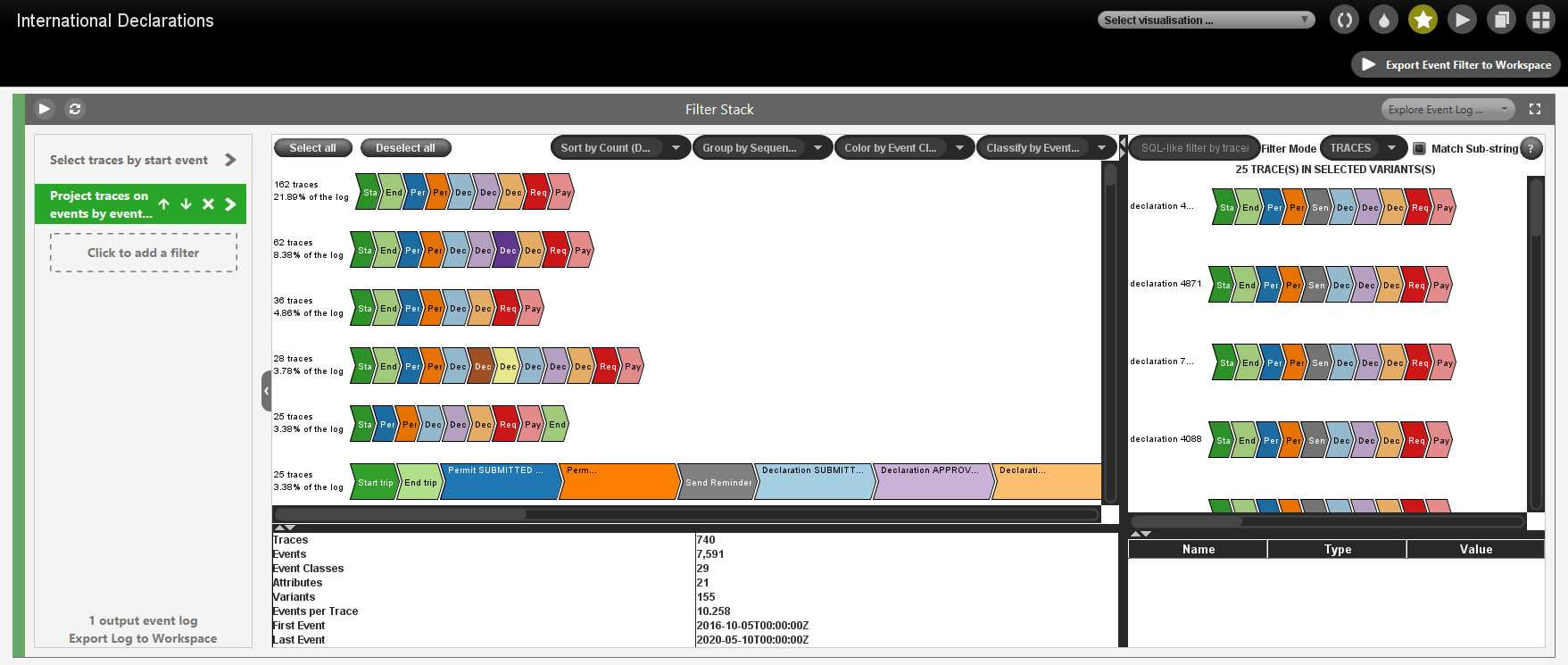}
  \caption{Interactive event log filtering in ProM, accessible via the ``Filter Event Log'' plugin. The shown event data has been filtered using two filters executed one after the other.}\label{fig:event:pre-processing:prom}
\end{figure}

As in any data analysis, identifying which event log pre-processing operations to apply for the analysis question at hand is an iterative process. Process mining software supports this iterative process by letting the analyst interactively build a \emph{stack} of filtering operations that can be modified and re-arranged alongside a visualization of the outcome of the filtering operation. Figure~\ref{fig:event:pre-processing:prom} shows the ``Filter Event Log'' plugin of ProM.

Other pre-processing operations on event data are
\begin{itemize}
\item Clustering the cases in the event log $L$ into multiple sub-logs $L_1,\ldots,L_k$ so that cases in a sub-log $L_i$ have similar trace variants and cases in different sub-logs $L_i \neq L_j$ have maximally different trace variants. Technically, clustering is repeated selection. However, the selection criteria are not based on selection predicates; see~\cite{DBLP:conf/icpm/ZandkarimiRSH20} for a survey on available clustering techniques.
\item Event data abstraction is a form of aggregation where arbitrary patterns in the event data are aggregated into higher level events; see~\cite{DBLP:journals/widm/DibaBWW20} for an overview.
\end{itemize}

\section{Event logs have limitations}\label{sec:event:log_limitations}

Event logs as defined in this document face severe limitations.
\begin{itemize}
\item The timestamp information in event data is often not reliable. For example, if events are only recorded on day-level granularity and three events $e_1,e_2,e_3$ occurred on the same day, then their order $e_1,e_2,e_3$ in the event table may not be the order in which they occurred. When creating an event log, we have to pick on ordering of these events to build a trace, but it may be the wrong one. A possible solution is to define a trace $\pi_{trace}(c)$ not as a sequence $\trace{e_1,\ldots,e_n}$ of events, but as a \emph{strict partial order} $(E,<)$ where $e_i < e_j$ are ordered only iff $\pi_{time}(e_i) < \pi_{time}(e_j)$. Events with the same time-stamp remain unordered.
\item Structured event logs only order event data according to a single case identifier. However, we have seen that even basic event data contains multiple entity identifiers that are in 1:n and n:m relationships to each other. For example in Table~\ref{tab:event:event_table:by_orderid}, customer A7001 is involved in 3 orders and deliver 623 is involved in 2 orders. The data structure of the structured event log cannot capture these relations. Graph-based data structures such as \cite{DBLP:books/sp/22/Fahland22,DBLP:journals/jodsn/EsserF21} allow tracing the behavior of multiple objects together.
\end{itemize}

\bibliographystyle{splncs04}
\bibliography{processmining_lit}

\end{document}